\documentclass[a4paper, 12pt]{article}
\usepackage{amsmath}
\begin{document}
\setlength{\baselineskip}{1.2\baselineskip}
\setlength{\parskip}{4ex}
\title{Solvability of eigenvalues in \textit{$j^n$} configurations}
\author{Igal Talmi\\Weizmann Institute of Science,\\Rehovot, Israel}
  \date{}
  \maketitle
   \begin{abstract}
Eigenvalues of eigenstates in $j^n$ configurations (\textit{n} identical nucleons in the \textit{j}-orbit) are functions of two-body energies. In some cases they are linear combinations of two-body energies whose coefficients are independent of the interaction and are rational non-negative numbers.
It is shown here that a state which is an eigenstate of \textit{any} two-body interaction has this \textit{solvability} property. This includes, in particular, any state with spin \textit{J} if there are no other states with this \textit{J} in the
$j^n$ configuration. It is also shown that eigenstates with solvable eigenvalues have definite seniority \textit{v} and thus, exhibit partial dynamical symmetry.

\end{abstract}
\section{Introduction and main results}

In this paper, we consider systems of \textit{n} identical fermions in a given
\textit{j}-orbit. Such $j^n$ configurations of protons or neutrons occur, for instance, in the nuclear shell model. Each configuration contains a certain number of fully antisymmetric independent states allowed by the Pauli principle. We consider Hamiltonians which contain the most general mutual two-body interactions between the particles. We consider interactions which are rotationally invariant and hence, eigenstates have definite values of angular momenta \textit{J}.

If there is only one state with given \textit{J} in the $j^n$ configuration, it is an eigenstate of any two-body (rotationally invariant) Hamiltonian. If there are several independent states with the same value of \textit{J}, they define a sub-matrix of the Hamiltonian whose diagonalization yields the eigenvalues as well as the eigenstates with the given \textit{J} for the mutual interaction considered. All matrix elements between states of the $j^n$ configuration are functions of matrix elements in the
$j^2$ configuration. The latter matrix elements are simply the energies in the two-particle configurations.

There are cases in which the expression of an eigenvalue is very simple. It is given by a linear combination of two-body energies with coefficients which are the same for all two-body interactions. These coefficients which are independent of the actual values of the
two-body energies, are rational numbers and the non-vanishing ones are positive. Such eigenvalues will be referred to as \textit{solvable}. An example may be seen in a recent paper [1], in which Zamick and Van Isacker consider certain states of the (9/2)$^4$ configuration whose
eigenvalues are solvable. They also calculate the energies of the \textit{J=2} and \textit{J=4} states of the (7/2)$^4$ configuration whose eigenvalues are solvable. In their arXiv version [arXiv:0803.1569vl(nucl-th] they "note that this derivation constitutes a proof that the coefficients...in the energy expression...must be rational numbers". The aim of the present paper is to determine in which cases these features, defined as \textit{solvability} in ref.[2], occur.

In $j^n$ configurations for $j<7/2$ any two-body interaction has dynamical symmetry [3] and is \textit{exactly solvable} [4]. Eigenvalues of all states in all $j^n$ configurations are given by linear combinations of two-body energies. The coefficients of those are rational functions of \textit{n}, \textit{J} and the seniority quantum number \textit{v}. For \textit{j=7/2}, eigenvalues of all states are solvable but are not given by a closed formula. For higher values of \textit{j}, this is no longer the case.

If there is only one state in the $j^n$ configuration with given value of \textit{J}, it is an eigenstate of \textit{any} (rotationally invariant) two-body interaction. It will be shown below that such a state has a solvable eigenvalue. Since there is only one such state, its structure must be independent of the interaction. If there are several states with the same value of \textit{J} they define a
sub-matrix of the Hamiltonian, characterized by \textit{n} and \textit{J}, which should be diagonalized. The eigenvalues are the roots of the secular equation whose degree is equal to the order of the sub-matrix. Apart form special cases, these eigenvalues are not linear in the two-body
energies. They are more complicated algebraic functions of the two-body energies. Still, there are cases in which eigenvalues are solvable also in such a situation. We will show that \textit{a state which is an eigenstate of any two-body interaction is solvable}. We will also show that if an eigenvalue is solvable, the corresponding state is an eigenstate of any two-body interaction.

Any state in the $j^n$ configuration may be expressed as
\begin{align}\Sigma[j^{n-2}(a_1J_1)j^2(J^\prime)J|\}j^n a J]\psi (j^{n-2}(a_1J_1)j^2_{n-1,n-2}(J^\prime)J)
\end{align}
where the coefficients are real and independent of the interaction. The states (1) are taken to be normalized, the sum of squares of the coefficients in (1) is equal to 1. The functions in the linear combination (1) are fully antisymmetric in particles 1 to \textit{n-2} and the $j^2$ wave functions are also antisymmetric. Each function in the sum, is not necessarily fully antisymmetric but their linear combination (1) is. In fact, all functions in (1) form an orthonormal basis for the space of all functions with the symmetry property described above. The fully antisymmetric $j^n$ functions belong to this space and hence, may be expanded in terms of the basis functions. The coefficients in (1) are called $n-2\rightarrow n$ coefficients of fractional parentage. They were introduced by Racah\cite {Racah} and their properties are described in detail in ref.[4]. The summation in (1) is over all values of $J_1$ and $J^\prime$ but instead of summing over all states with given $J_1$ characterized by $a_1$, it is always possible to use a basis in which only \textit{one} $j^{n-2}$ state with $J_1$ appears in (1).

Since the wave functions (1) are fully antisymmetric, the expectation value of a two-body interaction $\Sigma V(ik)$ may be calculated as the expectation value of one term, like $V(n-1,n-2)$, multiplied by the number of such terms, $n(n-1)/2$. The integration over the variables of particles 1 to $n-2$ may be carried out, leading to the expectation value
\begin{align}
(n(n-1)/2) \Sigma[j^{n-2}(a_1J_1)j^2(J^\prime)J|\}j^n a J]^2<j^2J^\prime|V|j^2J^\prime>
\end{align}
If the state (1) is an eigenstate of any two-body interaction, its eigenvalue is equal to its expectation value (2) and is thus, a linear combination of two-body energies. The coefficients of these are constants, independent of the actual values of the two-body energies.
These coefficients are squares of the expansion coefficients in (1) multiplied by $n(n-1)/2$ and hence, the non-vanishing ones are positive.
To show solvability, it should be proved that they are rational. This will be carried out in the next section.

The inverse statement is also true. A solvable eigenvalue is given by
\begin{align}
\sum c(J^\prime)<j^2J^\prime|V|j^2J^\prime>
\end{align}
with positive coefficients which are independent of the interaction. The corresponding eigenstate may be expressed by (1).  Since the \textit{V}(\textit{J}) are independent, equating (2) and (3) leads to the relation
\begin{align}
(n(n-1)/2) \sum[j^{n-2}(J_1)j^2(J^\prime)J|\}j^n a J]^2=c(J^\prime)
\end{align}
where the summation is over all values of $J_1$ and $a_1$. Due to the normalization of the c.f.p., the sum of squares of the $c(J^\prime)$ should be equal to $n(n-1)$/2.

To show that the coefficients in (3) are rational numbers we consider in the next section, the \textit{m}-scheme.

\section{The \textit{m}-scheme}

There is a simple way to calculate matrix elements of a two body interaction between states of \textit{n} fermions occupying a \textit{j}-orbit. It is based on the use of \textit{m}-scheme wave functions. Each wave function is defined by a set of ~~\textit{n}~ projections $m_1>m_2>...>m_n$. Such a wave function is a determinant in which the first row is  ~$\psi_{jm1}$(1),  $\psi_{jm1}$(2),...,  $\psi_{jm1}$(n). The second row is  $\psi_{jm2}$(1), $\psi_{jm2}$(2),...,  $\psi_{jm2}$(n) and so on. The exchange of variables of two particles is a change of two columns in the determinant which yields a change of sign. The normalization factor is $(n!)^{-1/2}$. The \textit{m}-scheme is described in detail in ref.[4] where matrix elements of a two-body interaction are derived. Diagonal matrix elements are equal to  $\Sigma<m_im_k|V|m_im_k>$ where the summation is over all pairs of \textit{m}-values in the state considered.

Matrix elements of a two-body interaction between \textit{m}-states
with given total $M=\Sigma m_i$ form a sub-matrix. Its trace is equal to the sum of diagonal elements taken between all states with given \textit{M}. States of the $j^n$ configuration with given values of
\textsl{J} and $M,J\geq M$, are linear combinations of these \textit{m}-states.
The linear transformation, from \textit{m}-states to states with definite \textit{J} and \textit{M}, is unitary or rather orthogonal.  Hence, the trace of the sub-matrix is invariant and is equal to the sum
 $\Sigma <j^naJ,M|V|j^na J,M>$ over all values of $J\geq M$ and over \textit{a} which characterizes different states with the same value of \textit{J}
if there are any. Hence, the trace may be used to calculate diagonal matrix elements of a rotationally invariant
\textit{M}-independent interaction in the \textit{J}-scheme.

The highest possible value of \textit{M} is reached in a state where $m_1=j, m_2=j-1,\ldots, m_n=j-n+1$. This value, $M_{max}$, is equal to
$nj-n(n-1)/2$. There is only one state with this $M_{max}$ and hence this state is the state with $J=M_{max}=nj-n(n-1)/2$. Its two-body interaction energy is that of the state with $M= M_{max}$  which is evaluated as explained above.

Subtracting the trace of the $M=J+1$ sub-matrix from the one of $M= J$, yields the sum   $\Sigma <j^na J,M|V|j^na J,M>$  over $J\geq M$ and over \textit{a} which
characterizes the states if there are several states with given \textit{J}  in the $j^n$ configuration. The number of states with \textit{J}= \textit{M} is equal to the number of states with \textit{M} from which the number of \textit{M}+1 states is subtracted. If they are equal, like for $M=M_{max}$-1, there is no state with \textit{J=M}. If there is only one such state, this procedure yields its interaction energy.
Otherwise, only the sum of energies of all states with given \textit{J}  is obtained. This case will be discussed in detail
in the following.

The diagonal matrix elements in the \textit{m}-scheme may be expressed in terms of matrix elements of two particles
coupled to a given \textit{J}  value. The two-particle matrix
elements in the \textit{m}-scheme may be transformed by using Clebsch-Gordan coefficients as follows.
\begin{align}
&<m_im_k|V|m_im_k>=\cr
&\Sigma <j,m_i,j,m_k|j,j,J,M=m_i+m_k>^2<j^2JM|V|j^2JM>
\end{align}
\noindent
Since the \textit{m}-states in (5) are antisymmetric, the sum on the r.h.s. of (5) is only over \textit{even} values of \textit{J}  (states of two fermions with \textit{odd} \textit{J}  values are \textit{symmetric}).

It follows from (5) that the coefficients of $V(J)=<j^2JM|V|j^2JM>$ in the diagonal elements in the \textit{m}-scheme are equal to sums of squares
of Clebsch-Gordan coefficients. A formula for these
coefficients was obtained by Racah and is given by (15.36), ref.\cite{Shalit},
\begin{align}
&(j,m_1,j,m_2|j,j,J,M)=[(j+m_1)(j-m_1)(j+m_2)(j-m_2)\times\cr
&(J+M)(J-M)]^{1/2}(2J+1)^{1/2}\Delta (jjJ)R(C-G)
\end{align}
where $R(C-G)$ is a rational function of the \textit{j}-values and
\textit{m}-values, and $\Delta (jjJ)$ is defined by
\begin{align}
\Delta(abc)=[(a+b-c)!(b+c-a)!(c+a-b)!/(a+b+c+1)!]^{1/2}
\end{align}	
It is seen that the squares of (6) are rational functions of their arguments. It is worth-while to point out that $m_1$ and $m_2$ appear under the square root sign only in the products of $(j+m_1)(j-m_1)$ and $(j+m_2)(j-m_2)$.

Thus, the sum  $\Sigma _{aJ}<j^na J,M|V|j^na J,M>$ is a linear combination of two-body matrix elements
$V(J)=<j^2JM|V|j^2JM>$. The coefficients of these combinations which do not vanish, are rational and positive numbers. If there is only one state with given \textit{J}  in the $j^n$ configuration, it is solvable. Its eigenvalue of a two-body interaction is a linear combination of expressions (5) of the $V(J)$  with positive rational coefficients. The state with $J=M_{max}$ considered above, is such a state.  If there are several states with the same \textit{J}, it was shown above that this property holds only for the sum of their diagonal elements.

An important property of the coefficients of the linear combinations is their independence of the two-body interaction. They are determined only by the properties of the Clebsch-Gordan coefficients and may be used for any two-body interaction. This is a feature of the \textit{m}-scheme. In fact, non-diagonal
matrix elements in the \textit{m}-scheme are also linear combinations of the $V(J)$. Two different \textit{m}-scheme
states which may have a non-vanishing matrix element between them must differ only by one pair of \textit{m}-values. This matrix element is equal to the matrix element between the different pairs of
\textit{m}-values, satisfying the relation $m_i+m_k = m_i^\prime +m_k^\prime =M,$
\begin{align}
&<m_im_k|V|m_i^\prime m_k ^\prime >=\cr
&\Sigma_J<j,m_i,j,m_k|j,j,J,M>\times\cr
&<j,m_i^\prime ,j,m_k^\prime |j,j,J,M><j^2JM|V|j^2JM>
\end{align}
The coefficients of the linear combination in (8) depend only on the values of the Clebsch-Gordan coefficients and are the same for any two-body interaction. From the properties of the Clebsch-Gordan coefficients (6), follows that the coefficients in (8) are all equal to a common square root which depends only on the \textit{m}-values (and \textit{j}) multiplied by rational functions of \textit{m}-values, the \textit{J}-value and \textit{j}.

Thus, diagonal matrix elements, as well as non-diagonal ones, are linear combinations of the two-body energies. The coefficients are determined by the Clebsch-Gordan coefficients and are the same for all two-body interactions. This scheme, mainly for several mixed configurations, has been amply used in large scale shell model calculations.

As explained above, states of the \textit{m}-scheme with given $M<M_{max}$-1 do not have definite values of \textit{J}. The matrix of the Hamiltonian constructed from them should be diagonalized to yield eigenstates with $J>M$ and the corresponding eigenvalues. An eigenstate is a linear combination of the various \textit{m}-states with coefficients $x_i$ where \textit{i}=1,...,\textit{N} and \textit{N} is the order of the matrix. There are always states which are eigenstates of any two-body interaction. These are states with $J>M$ which are the only states with given \textit{J} in the $j^n$ configuration considered.
There may be also other eigenstates with $J>M$ and also such eigenstates with $J=M$. Any eigenstate has a definite value of \textit{J} so we will consider conditions on any eigenstate of any two-body interaction.

We denote by $V_{ij}$ the elements of the sub-matrix characterized by \textit{M}. As explained above, the diagonal elements are linear combinations of the \textit{V}(\textit{J}) with rational coefficients. A non-diagonal element $V_{pq}$ is a product of square roots of rational functions $a_p$ and $a_q$ from the Clebsch-Gordan coefficients in (8),$a_p=[(j+m_i)(j-m_i)(j+m_k)(j-m_k)]^{1\over{2}}$ and $a_p=[(j+m_i^\prime)(j-m_i^\prime)(j+m_k^\prime)(j-m_k)]^{1\over{2}}$ and $R_{ij}$ which is a linear combination of the $V(J)$
 with rational coefficients. As pointed out above, $a_p$ and $a_q$ are independent of \textit{J}.

An eigenstate of the submatrix, with components $x_i$, should satisfy the conditions
\begin{align}
                        x_jV_{jj}+\Sigma_{x_i}V_{ji}= x_jV_{jj}+\Sigma x_ia_ia_jR_{ji}=Ex_j
\end{align}
Dividing this equation by a non-vanishing $x_j$, (9) may be expressed as
\begin{align}
          V_{jj}+\Sigma x_ia_iR_{ji}a_j/x_j=V_{jj}+\Sigma(x_ia_i/a_jx_j)R_{ji}a_j^2=E
\end{align}
Using these relations we obtain the following relations
\begin{align}
     V_{jj}+\Sigma(x_ia_i/a_jx_j)R_{ji}a_j^2 = V_{kk}+ \Sigma(x_ia_i/a_kx_k)R_{ki}a_k^2
\end{align}
By its definition, $a_i$ is the square root of a rational number and thus, the coefficients in (10) and (11), $R_{ki}a_k^2$ are all rational. If $x_j$ in (9) happens to be zero, a simpler relation follows. From (9)  follows then the relation
\begin{align}
 \Sigma x_ia_iR_{ji} =0
 \end{align}

Since the eigenstate considered should be an eigenstate of any
two-body interaction, every relation (11) is a set of several equations. They are obtained by putting in turn only one $V(J)$ equal to 1 and all others to zero. Each equation may be considered as an equation for the unknowns ($x_ia_i/a_kx_k$) which are the same for all \textit{J} values. From each relation (11), (2\textit{j}+1)/2 equations are obtained, which is the number of energies in the $j^2$ configuration. The coefficients of the unknowns are all rational numbers. Hence, the solutions of these linear equations are also rational. Similarly, from (12) follows that the unknowns $x_ia_i$ are rational numbers perhaps multiplied by a \textit{common}  irrational coefficient. Thus, the quotients ($x_ia_i/a_kx_k$) are rational. Using this fact in (10) we find that also E, the eigenvalue of the state considered, is a linear combination of two-body energies \textit{V}(\textit{J}) whose coefficients are not only non-negative but also rational.

There are solutions of the equations due to (10) whose number is at least equal to the number of states with $J>M$ which are eigenstates of any two-body interaction. If there is also such an eigenstate with $J=M$ there is a corresponding solution. It yields the coefficients of the linear combination of the \textit{V(J)} in the corresponding solvable eigenvalue. They are rational and the non-vanishing ones are positive.

The case of \textit{N}=2 is special since there is only one relation (10) and hence, it needs a special discussion. This case occurs only for
 $M=M_{max}$-2. One linear combination with coefficients \textit{x} and \textit{y}, of the two \textit{m}-scheme states is the $M=M_{max}$-2 component of the state with $J=M_{max}$. Its eigenvalue is given by $x^2V_1$+2$xyV_{12}+y^2V_2$ where \textit{x} and \textit{y} may be simply calculated from the $M=M_{max}$ component. Due to their values, every term in that expression is a linear combination of the \textit{V(J)} with rational coefficients. The orthogonal state which is the $M=M_{max}$-2 component of the $J=M_{max}$-2 state, is $y|1>-x|2>$. Its eigenvalue, given by $y^2V_1-2yxV_{12}+x^2V_2$, is also solvable in accordance with it being the only state with $J=M_{max}$-2.

Up to now, the discussion was about eigenstates of any two-body interaction without reference to their characteristics beyond their
spin \textit{J}. A very useful additional quantum number which may be used for a finer classification of states, is the seniority quantum number \textit{v}. In the next section, a brief description of the seniority scheme and its relevance to the subject of this paper will be presented.

\section{The seniority scheme}

The seniority quantum number \textit{v} measures in some sense, the amount of pairing of particles in \textit{J=0} states. It was introduced for atomic electrons in atomic spectroscopy by Racah\cite{Racah} and for nuclei, in
\textit{jj}-coupling, by Flowers[6] and Racah\cite{Racah1}. A detailed description of it and its applications may be found in refs.[4,8]. States of the seniority scheme are eigenstates of the pairing interaction, defined by its
two-body matrix elements
\begin{align}
<j^2JM|V|j^2JM>=(2j+1)\delta_{J0}\delta_{M0}
\end{align}
The eigenvalues of the pairing interaction  of states in the $j^n$ configuration with seniority \textit{v} are given by
\begin{align}
(n-v)(2j+3-n-v)
\end{align}

From (14) follows that in states with seniority \textit{v} in the
$j^v$ configuration there is no \textit{J=0} pairing. From such states, a chain of states with the same seniority \textit{v} and the same  \textit{J} may be obtained by multiplying them by states  of $(n-v)$/2 pairs coupled to \textit{J}=0 and antisymmetrizing. Thus, \textit{v} is the number of unpaired particles. The \textit{J}=0 state with no \textit{j}-particles has seniority \textit{v}=0. From it a chain of \textit{v}=0 and \textit{J}=0 states can be constructed in all $j^n$ configurations with \textit{n} even. The one particle state with \textit{J=j} has no pairing and \textit{v}=1. From it, a chain of \textit{v}=1,\textit{J}=\textit{j} states can be constructed in all $j^n$ configurations with \textit{n} odd. States with $J>0$, even, in the $j^2$ configuration have seniority \textit{v}=2. These examples are rather simple, but in general, seniority does not specify states uniquely. There are usually several independent states with the same value of \textit{J} and the same seniority \textit{v}.

There is a large class of two-body interactions which are diagonal in the seniority scheme. This class is far wider than the pairing interaction and the zero range  $\delta$-potential. Such interactions have vanishing matrix elements between states with different seniorities (and the same \textit{J}). Usually, however, such non-diagonal matrix elements do not vanish if the most general two-body interactions are considered. Still, there may be states with definite seniorities which are eigenstates of any two-body interaction. In fact, \textit{any eigenstate of any two-body interaction, which was shown above to have a solvable eigenvalue, has a definite seniority v}.

Let us first consider such eigenstates which are the only ones with given \textit{J} in the $j^n$ configuration. They cannot be admixtures of states with different seniorities like $x|vJ>+y|v^\prime J>$. Such a mixture implies the existence of \textit{two} independent, fully antisymmetric states, $|vJ>$ and $|v^\prime J>$ in contradiction to the eigenstate being the only state with
given \textit{J}. Also if there are several states with the same value of \textit{J},  admixtures of states with different seniorities cannot yield an eigenstate of \textit{any} two-body interaction. The amount of mixing is determined by the non-diagonal matrix elements. If we choose an interaction which is diagonal in seniority, like the pairing interaction, a mixture of states with different seniorities cannot be an eigenstate.

Clearly, matrix elements of two-body interactions which are diagonal in the seniority scheme have vanishing matrix elements between states with different seniorities. This is not the case for the most general two-body interactions. There are, however, certain selection rules which lead to vanishing of matrix elements between states with different seniorities of \textit{any} two-body interaction. These rules will be useful in finding states with definite seniorities which are eigenstates of any two-body interaction. A single particle operator acting on any state can change at most the state of one particle. Operating on a state with seniority \textit{v}, it can create a state with seniority \textit{v}, \textit{v}-2 and at most \textit{v}+2. It can complete a pair, reducing the seniority by 2, not change the "number of pairs" or destroy a pair, increasing the seniority by 2. Similarly, two-body operators may have
 non-vanishing matrix elements between states whose seniorities differ by 0, 2 or 4. Thus,
there are no non-vanishing matrix elements of any two-body interaction, between states with seniority \textit{v} and states with seniorities $v\pm k$ if $k$=6,8,...

Another selection rule applies to states in the middle of the shell, where \textit{n}=(2\textit{j}+1)/2. In such configurations, matrix elements of any two-body interaction vanish between states whose seniorities differ by 2. A simple example is offered by (7/2)$^n$ configurations, in which for $n<$4 or $n>$4 there is only one state with given \textit{J}. They all have definite seniorities and solvable eigenvalues. In the case \textit{n}=4 the states with \textit{J}=0(\textit{v}=0), \textit{J}=5(\textit{v}=4) and \textit{J}=8(\textit{v}=4) have solvable eigenvalues since they are the only ones with these spins. There are also two \textit{J}=2 and two \textit{J}=4 states with seniorities \textit{v}=2 and \textit{v}=4. Since \textit{n}=4 is the middle of the shell, there are vanishing matrix elements between \textit{v}=2 and \textit{v}=4 states. Thus, they are eigenstates of any two-body interaction and their eigenvalues are solvable.

As remarked in ref.[2], such cases are examples of partial dynamical symmetry [9,10]. States are characterized by the seniority which is a symmetry associated with the SU(2) Lie group, even though the Hamiltonian of which they are eigenstates does not have this symmetry.  In the (9/2)$^5$ configuration, states with seniorities \textit{v}=3 and \textit{v}=5 cannot mix.  Hence, the states with $v=3$ and $v=5$ and spins \textit{J}=5/2, \textit{J}=7/2, \textit{J}=11/2, \textit{J}=13/2, \textit{J}=15/2, and \textit{J}=17/2 are eigenstates of any two-body interaction. Also the \textit{J}=9/2 state with \textit{v}=3 cannot mix with the \textit{v}=1 and \textit{v}=5 states and hence, is an eigenstate. This is unlike the situation in the (9/2)$^3$ configuration.  In any $j^3$ configuration for $j>7/2$, there are non-vanishing matrix elements of the most general two-body interaction between the $J=j, v=1$ state and all $J=j, v=3$ states.  In fact, all non-vanishing matrix elements between $j^n$ states with different seniorities are linear combinations of these $j^3$ matrix elements \cite{Shalit}. Other states with definite seniorities in this configuration are those which are the only ones with given \textit{J} in the $(9/2)^5$ configuration. These are the \textit{J}=1/2, \textit{J}=19/2 and \textit{J}=25/2 with seniorities \textit{v}=5 and the \textit{J}=3/2, \textit{J}=21/2 with \textit{v}=3. All these states are solvable.

Other examples of this kind are in the $(11/2)^6$ configuration, where there are $J=0$ states with seniorities 0, 4 and 6. The \textit{v}=6 state cannot have a non-vanishing matrix element with the \textit{v}=0 state. There is no such non-diagonal matrix element between the \textit{v}=6 and \textit{v}=4 state since the configuration is in the middle of the \textit{j}=11/2 orbit. Thus,
the state with \textit{J}=0, \textit{v}=6 is an eigenstate of any two-body interaction and its eigenvalue is solvable.  Other eigenstates of this kind in this configuration are  the \textit{J}=3, \textit{v}=4 state which cannot mix with the two \textit{J}=3, \textit{v}=6 states. Also the \textit{J}=5, \textit{v}=6 state (no mixing with the two \textit{v}=4 states), and the \textit{v}=4 and \textit{v}=6 states with \textit{J}=11, \textit{J}=13 and \textit{J}=14 are eigenstates of any two-body interaction. They have definite seniorities and are solvable.

In this configuration there are other states with definite seniorities. There are two \textit{J}=2, \textit{v}=4 states which, as explained above, have vanishing matrix elements with the \textit{J}=2,\textit{v}=2 and \textit{J}=2,\textit{v}=6 states. Still, it is not clear whether  there are linear combinations of them which are eigenstates of any two-body interaction. A similar situation occurs for the two \textit{J}=3,\textit{v}=6 states, the 3 states with \textit{v}=4 and \textit{J}=4, \textit{J}=6 and \textit{J}=8 and several other states.
The states described above have definite seniorities and yet it is not known whether any linear combinations are solvable. In any case they exhibit partial dynamical symmetry.

In all cases mentioned above, solvability follows from the selection rules. There may be, however, other solvable eigenvalues which do not follow from the general rules. As long as their origin is not known, they may be called "accidental". Such a case is described in ref.[1]. In the (9/2)$^4$ configuration, there are 3 states with \textit{J}=4 and 3 states with \textit{J}=6. One of the \textit{J}=4 (and one of the \textit{J}=6) states may be taken to be the \textit{v}=2 state and the two states orthogonal to it have seniority \textit{v}=4. In an earlier paper, Zamick et al[11] found that one \textit{v}=4 state may be constructed so that it has vanishing matrix elements with the \textit{v}=2 state and also with the orthogonal \textit{v}=4 state for any two-body interaction. Thus, the special state with \textit{J}=4 (and the one with \textit{J}=6) is an eigenstate of any two-body interaction and hence, has a solvable eigenvalue. It is a linear combination of two-body energies \textit{V(J)} whose coefficients are rational and non-negative as actually calculated in ref.[1]. These special \textit{J}=4 and \textit{J}=6 states have definite seniority \textit{v}=4 unlike the other \textit{v}=4 states which have non-vanishing matrix elements of a general two-body interaction with the \textit{v}=2 states. Also here is a case of partial dynamical symmetry.

\section{Summary}

The paper considers eigenstates of Hamiltonians with single particle energies and with the most general two-body interactions, whose eigenvalues are solvable. Within $j^n$ configurations, they are expressed as linear combinations of two-body energies $V(J)=<j^2J|V|j^2J>$ whose coefficients are rational and non-negative. It is proved that \textit{any state which is an eigenstate of any two-body interaction has a solvable eigenvalue}. The inverse is also true. Any solvable eigenvalue belongs to a state which is an eigenstate of any two-body interaction. In particular, a state with given value of spin \textit{J} which is the only state in the $j^n$ configuration with this value of \textit{J}, is clearly an eigenstate of any two-body interaction. Hence, its eigenvalue is solvable.

The seniority scheme is closely related to solvable cases. It is shown that \textit{an eigenstate of any two-body interaction must have a definite seniority v}. There are selection rules on matrix elements of any
two-body interaction between states of the seniority scheme. These help in finding examples of solvable cases, some of which are listed.

The most general two-body interaction is not diagonal in the seniority scheme. Still, there are eigenstates of such interactions and they have definite seniorities. These are examples of partial dynamical symmetry. Some eigenstates carry quantum numbers \textit{v} which characterize irreducible representations of the SU(2) group, whereas the Hamiltonian does not possess this symmetry. It cannot be constructed only from odd tensor and \textit{k}=0 tensor operators which commute with the generators of SU(2).


\begin{thebibliography}{99}
\bibitem{Zamick}L.Zamick and P.Van Isacker, Phys. Rev.C \textbf{78} (2008) 044327.
\bibitem{Isacker}P.Van Isacker and S.Heinze, Phys. Rev. Lett. \textbf{100} (2008) 052501.
\bibitem{Iachello}F.Iachello, \textit{Lie Algebras and applications, Lecture Notes in Physics}, Springer Berlin and Heidelberg (2006), p.164
\bibitem{Shalit}A.de-Shalit and I.Talmi, \textit{Nuclear Shell Theory},
Academic Press (1963);Reprinted by Dover Publications (2003).
\bibitem{Racah}G.Racah, Phys. Rev. \textbf{63} (1943) 367.
\bibitem{Flowers}B.H.Flowers, Proc. Roy. Soc. (London) \textbf{A212}.
(1952) 248.
\bibitem{Racah1}G.Racah, in \textit{L.Farkas Memorial Volume}, Research Council of Israel (1952) 294.
\bibitem{Talmi}I.Talmi, \textit{Simple Models of Complex Nuclei}, Harwood Academic Publishers (1993).
\bibitem{Alhassid}Y.Alhassid and A. Leviatan, J. Phys. A \textbf{25}(1992) L 1265.
\bibitem{Leviatan}A.Leviatan, Phys. Rev. Lett. \textbf{77} (1996) 818.
\bibitem{Ascuderos}A.Ascuderos and L.Zamick, Phys. Rev. C \textbf{73} (2006) 044302.
\end{thebibliography}
\end{document}